\documentclass[12pt,preprint,colordvi,epsf]{aastex}

\newcommand{\msun}{$M_{\odot}$}

\newcommand{\rp}{$r$-process}

\newcommand{\feh}{[Fe/H]} 

\newcommand{\rfe}{[$r$/Fe]}
\newcommand{\cfe}{[C/Fe]}
\newcommand{\nfe}{[N/Fe]}
\newcommand{\ofe}{[O/Fe]}
\def\h2{H$_{2}$}

\input epsf

\begin{document}

\title{Nucleosynthesis, Reionization, and the Mass Function of the First Stars}
\author{
JASON TUMLINSON\altaffilmark{1},
APARNA VENKATESAN\altaffilmark{2,3} \& J. MICHAEL SHULL\altaffilmark{2,4}}
\altaffiltext{1}{Department of Astronomy and Astrophysics,
                 University of Chicago,
                 5640 S. Ellis Ave., Chicago, IL 60637}
\altaffiltext{2}{CASA, UCB 389, Department of Astrophysical and Planetary
  Sciences, University of Colorado, Boulder, CO 80309-0389}
\altaffiltext{3}{NSF Astronomy and Astrophysics Postdoctoral Fellow}
\altaffiltext{4}{Also at JILA, University of Colorado and National Institute of
        Standards and Technology}

\begin{abstract}
We critique the hypothesis that the first stars were very massive stars
(VMS; $M > 140$~\msun). We review the two major lines of evidence for
the existence of VMS: (1) that the relative metal abundances of extremely
metal-poor Galactic halo stars show evidence of VMS enrichment, and (2)
that the high electron-scattering optical depth ($\tau_e$) to the CMB
found by {\it WMAP}  requires VMS for reionization in a concordance
$\Lambda$CDM cosmology.  The yield patterns of VMS exploding as
pair-instability supernovae are incompatible with the Fe-peak and
$r$-process abundances in halo stars.  Models including Type II
supernovae and/or ``hypernovae'' from zero-metallicity progenitors
with $M = 8 - 40$~\msun\ can better explain the observed trends.
We use the nucleosynthesis results and stellar evolution models to
construct an initial mass function (IMF) for reionization. With a
simple metal-transport model, we estimate that halo enrichment curtails
metal-free star formation after $\sim 10^8$ yr at $z \sim 20$. Because
the lifetime-integrated ionizing photon efficiency of metal-free stars
peaks at $\sim$ 120~\msun\ and declines at higher mass, an IMF with
an approximate lower bound at $M \sim 10 - 20$~\msun\ and no VMS can
maximize the ionizing photon budget and still be consistent with the
nucleosynthetic evidence. An IMF devoid of low-mass stars is justified
independently by models of the formation of primordial stars. Using a
semi-analytic model for H~I and He~II reionization, we find that such an
IMF can reproduce $\tau_e \simeq 0.10 - 0.14$, consistent with the range
from {\it WMAP}, without extreme astrophysical assumptions, provided that
metal-free star formation persists $10^7 - 10^8$ yr after star formation
begins.  Because stars in the mass range 50 -- 140~\msun\ are the most
efficient sources of ionizing photons, but are expected to collapse to
black holes without releasing metals, this IMF effectively decouples
early metal enrichment and early ionization.  Such an IMF may allow the
unique properties of the zero-metallicity IMF to persist longer than they
would in the pure VMS case, and to contribute significantly to the global
ionizing photon budget before halo self-enrichment and/or inter-halo
metal transport truncates metal-free star formation. We conclude, on the
basis of these results, that VMS are not necessary to meet the existing
constraints commonly taken to motivate them.  \end{abstract}

\section{INTRODUCTION}

Theoretical ideas about the first, metal-free stars have developed
rapidly in recent years, despite the absence of direct observational
evidence of their properties. As a specific prediction of the big bang
theory, metal-free first stars confront both cosmology and astrophysics
generally with important questions: When did the first stars form? What
was their initial mass function (IMF)? What were their lifetimes and
feedback effects? Where are their remnants (compact objects and/or
metal enrichment)?  The standard tools of stellar evolution theory have
been applied to calculate the evolving structure and radiative spectra
of metal-free stars, given their mass (Tumlinson \& Shull 2000; Marigo
et al. 2000; Bromm, Kudritzki, \& Loeb 2001; Schaerer 2002; Tumlinson,
Shull, \& Venkatesan 2003, hereafter TSV03). Many uncertainties remain
about the cosmological context and feedback effects of the first stars,
the largest of which concern the duration of the metal-free phase and
the IMF.  Despite these unknowns, theoretical progress on the first
stars has arrived at six points of consensus:

\begin{itemize}
\item[1.] The first stars form from metal-free gas in the first collapsed
       dark matter halos around $z \sim 20$ (Ricotti, Gnedin, \& Shull
        2002a,b).
\item[2.] The first stars may be very massive ($M > 100$~\msun), because
       they are restricted during their formation to inefficient
       cooling by molecular hydrogen below $10^4$ K (Abel, Bryan, \&
       Norman 2000).
\item[3.] These very massive first stars seed their own halos and possibly
       enrich nearby ones by releasing metals from pair-instability
       supernovae (140 -- 260~\msun; Heger \& Woosley 2002, hereafter
       HW02).
\item[4.] At critical metallicity $Z \gtrsim 10^{-3.5} Z_{\odot}$,
       protostellar clouds are able to cool and fragment more
       efficiently, leading to a ``normal'' IMF (Bromm \& Loeb 2003;
       Schneider et al.~2002). Where this metallicity has not been
       reached, the IMF retains its unusual properties. Halo and IGM
       enrichment probably occur inhomogeneously in time and space
       during the transition to ``normal'' star formation.
\item[5.] The first stars begin, and may complete the reionization of
       intergalactic H~I (Gnedin \& Ostriker 1997) and He~II
       (Venkatesan, Tumlinson, \& Shull 2003, hereafter VTS03).  The
       total first-stars contribution to the global ionizing photon
       budget is unknown, but the efficient ionization of VMS is
       thought to be required. There may be partial recombination
       during the transition phase in metallicity, followed by a second
       reionization near $z_r = 6$ (Cen 2003a; Wyithe \& Loeb 2003; VTS03).
\item[6.] The first-stars epoch ends when all star forming regions
       have achieved the critical metallicity.  This is thought to occur
       before the final stages of H~I reionization at $z_r = 6.2$.
\end{itemize}

This consensus is not universally accepted in all details, but it
serves as a working hypothesis in the literature.  Point (2), the {\em
VMS hypothesis}, has appeared often in studies of the first stars
concerned with their cosmic origins, their sites of formation, and
their importance for reionization. In this paper, we examine this
hypothesis with the objective of constructing an IMF for the first
stars that is consistent with current observational constraints. We
meet this goal by examining the arguments for the VMS hypothesis and
by investigating how its key lines of evidence can also support an IMF
without VMS.

Our critique of the VMS hypothesis addresses consensus points (3) and
(5) particularly.  The empirical basis of point (3) is the work of
Oh et al. (2001) and Qian \& Wasserburg (2002), who argued that the
relative metal abundances in extremely metal-poor (EMP) halo stars
match the expected signatures from VMS (HW02). Their conclusion has
since been adopted as positive evidence of VMS metal-enrichment and
ionizing photon production in the first stellar generation. We argue
that this line of evidence is not supported by the broader sets of
abundance data presently available. Instead, we adopt the view of Umeda
\& Nomoto (2004, hereafter UN04) that these observed abundances are
better matched by core-collapse supernovae or ``hypernovae'' from 10 --
50~\msun\ progenitors. We then examine point (5) in light of evolving
spectra models for metal-free stars (TSV03) and recent studies of star
formation at zero metallicity. VMS have been suggested as the most viable
cause of the high Thomson electron-scattering optical depth $\tau _e
= 0.17^{+0.08}_{-0.07}$ derived from the cosmic microwave background
(CMB) data obtained by the Wilkinson Microwave Anisotropy Probe ({\it
WMAP}; Spergel et al.~2003).  Instead, we find that reionization models
incorporating an IMF that excludes both low-mass stars and VMS can
meet the reionization constraint.  The lower but equally uncertain
value $\tau _e = 0.12^{+0.08}_{-0.06}$ found by joint analysis of {\it
WMAP} and the SDSS matter power spectrum by Tegmark et al. (2004) is
reproduced by our models, but since it is also easier to accommodate
to a standard IMF we conclude that the optical depth to reionization
is not a decisive constraint on the IMF at early times. Because the
ionizing efficiency of metal-free stars increases more rapidly from 1
-- 50~\msun\ than it does above this range (see \S~4), we are able to
construct an IMF with ionizing properties similar to those of VMS. We
conclude, on the basis of these results, that VMS are not necessary to
meet the existing constraints commonly taken to motivate them.  We also
do not need to invoke high values for the escape fraction of ionizing
radiation or the star formation efficiency in halos (cf., Oh et al.
2001; Somerville, Bullock \& Livio 2003).

This paper is structured as follows. First, in \S~2, we review the
theoretical motivation for VMS in the first stellar generation and
distinguish two versions of the VMS hypothesis. In \S~3 we review
the extremely metal-poor halo star abundances and compare specific
patterns in the nucleosynthetic evidence to theoretical yields. In
\S~4 we use our results to motivate a range of possible first-stars
IMFs and derive a timescale for the epoch of metal-free star formation.
In \S~5 we discuss semi-analytic reionization models that compare these
different IMFs to VMS.  In \S~6 we draw general conclusions and discuss
some unsolved problems in the study of the first stars.

\section{The VMS Hypothesis}

We define VMS to include stars with $M \geq 140$~\msun, the lowest
mass at which metal-free stars explode as pair-instability supernovae
(PISN; HW02). This is the only clear mass cutoff higher than the uncertain
$M \sim 40 - 70$~\msun\ division between core-collapse SN and complete
collapse to black holes (HW02).  The $M \geq 140$~\msun\ definition
cleanly separates two mass regimes with very different nucleosynthetic
signatures (see \S~3) but similar radiative properties.  The mass cutoff
is rigid for the purposes of definition only; it may be more loosely
defined in reality without interfering with our conclusions, provided it
marks a qualitative change in the nucleosynthetic signature of the stars.

The theoretical motivation for the VMS hypothesis rests primarily on
the microscopic processes of radiative cooling in low-density
primordial gas. At $T \lesssim 10^4$~K, cooling by H and He is
negligible and metal-line cooling is dominated by [C~II], [O~I], and
[Si II] fine-structure lines. Because metal-free gas is restricted to
relatively inefficient cooling by H$_2$ at $T \lesssim 10^4$~K, cooling
rates are reduced, and the balance between pressure support and
gravitational collapse is shifted to higher temperatures.  Primordial
protostellar objects, being unable to cool or fragment, must therefore
be more massive to overcome their elevated gas pressure. The final star
should also be more massive if all the cloud can accrete.  This simple
physical description was rigorously implemented by Abel, Bryan, \&
Norman (2000), who used an adaptive mesh refinement hydrodynamics code
to follow the evolution of metal-free gas from cosmological initial
conditions down to the formation of a self-gravitating object with $M
\sim 600$~\msun. The simulation was halted when the core became
optically thick to \h2\ line emission, which occurred before the object
was completely formed. This study implies VMS with $M > 140$~\msun, if
all the gravitationally bound gas can accrete to a single star.

Recent studies (Omukai \& Palla 2002; Tan \& McKee 2002; Bromm \&
Loeb 2004), taking the Abel et al. (2000) result as a starting point
for semi-analytic models of primordial star formation, indicate that
reality is complicated by stellar feedback on the accreting matter.
Omukai \& Palla (2002) identified a critical accretion rate ($\dot{M}$
$ = 4 \times 10^{-3}$~\msun\ yr$^{-1}$) below which laminar accretion
onto a spherically-symmetric protostar can proceed to $M > 100$~\msun\
unimpeded by feedback.  Above this critical rate, the accretion luminosity
exceeds the Eddington limit and the flow is reversed at 100 -- 300~\msun.
Models by Bromm \& Loeb (2004) have found a conservative upper
limit of 500 \msun, by setting the time available to accretion to be the
stellar lifetime ($\sim 3 \times 10^6$ yr) but ignoring feedback. Tan \&
McKee (2002) consider the effects of disk accretion geometry, rotation,
and radiation feedback in limiting the mass of primordial stars. They
find that these feedback mechanisms are likely to operate at $M = 30 -
100$~\msun, perhaps limiting the masses of metal-free stars to this range.
However, they later found (Tan \& McKee 2004) that none of the feedback
mechanisms considered can halt the accretion before the star achieves
$\sim 30$ \msun.  This important result indicates that both VMS and
low-mass stars may not form at zero metallicity and suggests that the
mass limits of the primordial IMF may be quite different from today. In
\S~4 we use this idea and empirical constraints to construct an IMF for
the first stars.

These theoretical studies of primordial protogalaxies and star
formation posit that the first generation of stars was partially, and
perhaps completely, composed of VMS.  To capture existing evidential
claims, we introduce a distinction between the {\em strong} VMS
hypothesis (``the first generation were exclusively VMS'') and the {\em
weak} VMS hypothesis (``the first generation included VMS in addition
to $M \lesssim 50$~\msun\ stars'').  Both versions appear in the
literature, although not with these names.  The evidence can bear
differently on the two versions of the hypothesis, because the ionizing
properties and nucleosynthetic yields vary across the mass function.
These labels do not connote the relative merit of the two versions;
they suggest only that the ``strong'' hypothesis is more readily
falsified than the ``weak''. The ``weak'' hypothesis is expected to be
more difficult to constrain, because it includes low-mass stars and so
is more like the present-day circumstances. We evaluate these versions
of the hypothesis below.

The HW02 stellar evolution and supernova models show that, between
140 -- 260~\msun, zero-metallicity stars are highly susceptible
to the pair-production instability after core He depletion.  This
instability quickly disrupts the star, ejects metals, and leaves
no compact remnant.  Because it is triggered at an unusually early
period in the star's evolution, this pair-instability supernova (PISN)
produces an unusual nucleosynthetic signature.  If the VMS hypothesis is
correct, the signature of PISN could appear in the metal abundances of
``second-generation'' EMPs in the Galactic halo (see \S~3). Qian \&
Wasserburg (2002) and Oh et al. (2001) used this idea to argue that
these stars justify the VMS hypothesis.

Models with VMS have also been proposed to explain the current data on
IGM reionization, specifically the high electron-scattering optical
depth $\tau_e$ found in the {\it WMAP} CMB data.  For the simplest
possible reionization history (a step function at fixed $z_r$), this
result implies that complete reionization occurred at $z_r = 17 \pm 4$
(Spergel et al. 2003; Kogut et al. 2003).  Recent work that combines
the three-dimensional matter power spectrum, $P(k)$, from the Sloan Digital
Sky Survey with {\it WMAP} data finds that $\tau _e =
0.12^{+0.08}_{-0.06}$ (Tegmark et al.~2003). The large
uncertainties arise from the fact that the first-year {\it WMAP}
$\tau_e$ results were based on temperature-polarization correlations,
which are an indirect measure of $\tau_e$. Additional {\it WMAP} data
will allow auto-correlation of the polarized CMB maps and should
provide a direct and more precise determination of $\tau _e$.

When joined with spectroscopic observations that indicate a rapidly
rising H~I Gunn-Peterson opacity up to $z \sim$~6.4 (Becker et al.
2001; Djorgovski et al.  2001), the {\it WMAP} result implies that the
IGM experienced a relatively complex ionization history, with extended
periods of ionization in H~I and/or He~II at $z \sim$ 6 -- 20 (VTS03;
Cen 2003a; Wyithe \& Loeb 2003).  In a study of the ionizing properties
of VMS, Bromm, Kudritzki, \& Loeb (2001) found VMS with $M = 100 -
1000$~\msun\ to be an order of magnitude more efficient, {\it on the
zero-age main sequence}, at generating ionizing photons per unit
stellar mass than are stars in a normal IMF with $M = 1 - 100$~\msun\
(Tumlinson \& Shull 2000).  VMS produce efficient ionization and have
short lifetimes, and they are thought to form in the metal-free gas of
collapsed halos at high $z$. In light of these features, scenarios
involving VMS in combination with subsequent stellar populations have
been widely proposed as an attractive explanation for the observed
ionization and thermal history of the IGM (Hui \& Haiman 2003) and the
large detected $\tau_e$ to the CMB (Cen 2003b; Haiman \& Holder 2003;
Fukugita \& Kawasaki 2003; Ciardi, Ferrara \& White 2003; Kaplinghat et
al. 2003).

These ``nucleosynthesis'' and ``reionization'' lines of evidence are often
taken to support the VMS hypothesis. But are VMS required? Are there other
viable models for the EMP metal abundances and reionization that do not
include VMS? We argue that neither line of evidence is compelling. We
take up the nucleosynthesis arguments first.

\section{The First Stars and Nucleosynthesis}

\subsection{General Patterns in the Nucleosynthetic Evidence}

Many studies of reionization by the first stars (\S~2) have adopted the
VMS hypothesis, citing as evidence the apparent match between PISN
yields and the metal abundances of EMP halo stars (Oh et al. 2001; Qian
\& Wasserburg 2002).  A full review of the literature on metal
abundances (elements Li to U) in EMP stars is beyond the scope of this
study.  The review by McWilliam (1997) gives a survey up to that time,
and papers that cite it (available from the Astrophysical Data Service)
provide useful entry points to the more recent literature. We focus
here on three clear features in Galactic halo EMP abundances which are
confirmed by multiple studies and which bear directly on the VMS
hypothesis.  These are the specific Fe-peak element ratios
(especially Zn), the widespread presence of $r$-process elements, and
elevated [C,N,O/Fe] ratios.  We summarize these features here, and in
\S~3.2 we compare them in detail to theoretical models incorporating
the VMS hypothesis.

{\em Iron-peak elements (Cr -- Zn):} Because these elements are
produced in explosive events, they are sensitive to the energy,
rotation, mass cut, and asymmetry of the supernova and to the pre-SN
stellar properties. Large samples presented by McWilliam et al. (1995)
and Carretta et al. (2002) found that the Fe-peak ratios to Fe
qualitatively change their behavior at \feh\ $ \simeq -3$. Below this
metallicity, [Zn/Fe] and [Co/Fe] increase while [Cr/Fe] and [Mn/Fe]
decrease.  This trend was interpreted as a change in the nature of the
supernovae progenitors, and the increased scatter was attributed to
inhomogeneous enrichment. The more recent and higher resolution study by
Cayrel et al.~(2003) found lower scatter and a smooth rather than abrupt
change in the abundance ratios at \feh\ $\simeq -3$ to -4. In any case,
the overall abundance pattern is roughly consistent across the three
studies, and it provides a good test for theoretical yields.

{\em $r$-Process elements (A $>$ 100):} These elements are thought to
be produced by rapid neutron capture in hot, dense, neutron-rich
environments during explosive events. Although the exact physical sites
are still uncertain, the proposed mechanisms are all associated with
massive stars in the range $M = 8 - 40$~\msun\ (Truran et al. 2002).
The absolute abundances and relative ratios of $r$-process elements are
thus sensitive indicators of core-collapse supernova activity.  The
existing samples of EMP stars show widespread evidence of $r$-process
elements down to \feh\ $\sim -3$ (McWilliam et al. 1995; Burris et al.
2000). The mean [$r$/Fe] is similar to the solar value at all \feh, but
with up to 2 dex scatter at \feh\ $\sim -3$. The relative abundances
(i.e., [Eu/Ba]) are also similar to the solar values.

{\em Primary elements (C, N, O):} Of these direct products of
main-sequence stellar nucleosynthesis, C is easily found in EMPs, but N
and O are difficult to measure.  Many EMP stars are C-rich relative to Fe.
Further study of this class of stars may ultimately reveal them
to be {\em iron-poor}, rather than {\em metal-poor}, and therefore less
chemically primitive than their low \feh\ makes them appear.  As pointed
out by Bromm \& Loeb (2003), C and not Fe is the dominant interstellar
coolant that may influence the early IMF. As shown in TSV03, C is the
nuclear burning catalyst that determines by its absence the unusual
behavior of the first stars. These factors favor C instead of Fe as the
``reference element'' for finding chemically primitive stars in the local
universe. By this measure, these C-enhanced stars may be less primitive
than those with higher \feh.  For now however, we follow the observational
studies by assuming they are in same class as the other EMPs. 

\subsection{VMS and Nucleosynthesis}

In this section, we describe the nucleosynthetic evidence for VMS as
stated in the literature.  Observational studies of EMP metal abundances
have regarded the trends in the Fe-peak and \rp\ elements as the most
critical constraints on the nature of the first stars.  We examine
the model of Wasserburg \& Qian (2000) and Qian \& Wasserburg (2002),
which represents only one example from a broad class of possible models
that could explain the observed variations. We cannot address all the
possible variations in such a class, so we use this one because it admits
VMS objects as a possible source of early enrichment. We later use the
phenomenological model of Fields, Truran, \& Cowan (2002) to illustrate
why a successful model for the observed trends in \rp\ abundances has
difficulty incorporating VMS.

Wasserburg \& Qian (2000; hereafter WQ00) examined the behavior of \rfe\
with \feh\ in the McWilliam et al. (1995) sample and argued that the
qualitative change in \rfe\ at \feh\ $\lesssim -3$ presented an ``iron
conundrum''; namely, that the wide dispersion in Eu and Ba abundances at
\feh\ $ \simeq -3$ suggested unrelated sources of Fe and $r$ elements. In
response, they proposed a ``prompt'' (P) inventory of Fe production
by an initial population with large Fe yields but little or no $r$
production. This initial inventory ceased at \feh\ $\sim - 3$, where
the onset of high-frequency (HF) events (SN II; $\tau \sim 10^7$ yr)
and low-frequency (LF) events (SN Ia; $\tau \sim 10^8$) initiated
a correlation between $r$ and Fe\footnote{To avoid confusion with
hydrogen, we adopt HF and LF to designate the H and L events posited by
WQ00.}. In this model, the wide variation in Fe for small barium abundance
$\epsilon$(Ba) is attributed to diverse Fe yields for P events or to
multiple P progenitors.  The low $\epsilon$(Ba) at \feh\ $\lesssim -3$
is attributed to small co-production of $r$ elements by P events or to
dilution of a modest HF contribution.  WQ00 made no specific claims about
the mass or metallicity of the P inventory sources.

After establishing the general properties of the P inventory in WQ00,
Qian \& Wasserburg (2001) and Qian \& Wasserburg (2002; hereafter QW02)
noted that VMS of a few hundred~\msun\ possess the Fe-rich, $r$-poor
yields and short lifetimes ($1 - 3$ Myr) necessary to produce the P
inventory Fe over the short time ($\sim 10^7$ yr) before the onset of HF
events.  This VMS model readily accommodates the qualitative change in
observed abundances at \feh\ $ = -3$, near the $Z = 10^{-3.5}
Z_{\odot}$ threshold where metal-line cooling is expected to change the
IMF (Bromm \& Loeb 2003). This model was extended by Oh et al. (2001)
to link the VMS hypothesis to reionization.  They pointed out that the
trends in \feh\ and Ba also appear in Si and Ca, elements produced by
PISN, but not in C, which is not abundantly produced by PISN.  Their
conclusion has since been adopted as positive evidence that VMS
dominated metal enrichment from the first stellar generation.  On this
assumption, Oh et al. (2001) argued that the Fe enrichment in the P
inventory (\feh\ $= -3$) implies cumulative cosmic star formation
activity sufficient to reionize the universe if all star formation is
in VMS. This result forms the empirical basis of the contention that
VMS could explain both the observed EMP metal abundances and
reionization together (see points 3 and 5 in the introduction).

However, closer scrutiny shows that the P inventory does not match all
the observations.  The P-inventory model is incompatible with the strong
VMS hypothesis, because VMS have no significant post-He nuclear burning
and therefore produce no $r$ elements (HW02).  If VMS produce all the
Fe up to [Fe/H] $\sim -3$, the $r$ elements should be absent instead of
appearing at [$r$/Fe] $\gtrsim -0.5$ as observed.  Progenitors with $M =
8 - 40$~\msun\ are needed in the first generation to provide the observed
\rp\ elements (Truran et al. 2002), provided that the \rp\ mechanisms
that operate at finite metallicity can be safely extended to $Z = 0$.
Thus, the strong VMS hypothesis is convincingly excluded, and the weak
hypothesis is more naturally accommodated by the P-inventory model.

The P-inventory model also provides a means for testing the weak VMS
hypothesis, because it attributes virtually all the Fe to VMS. To see
why, we follow WQ00 and estimate the fraction of Fe in the P inventory
that could be attributed to HF events.  If we take the lowest
$\epsilon$(Ba) to represent the contribution of a single HF event, then
40 such events contribute to the scatter in $\epsilon$(Ba) at \feh\ $
\sim -3$ for a modest increase in \feh.  This small gain in Fe
enrichment limits the HF contribution to the P inventory to
$\lesssim10$\%.  Another approach to this proof comes from
consideration of a model IMF. If we assume a Salpeter IMF with $1 -
260$~\msun, adopt HW02 yields for VMS and a uniform Fe yield per star,
$m_{\rm Fe} = 0.07$~\msun\ for $M = 10 - 40$ \msun, we find that the
VMS mass range produces $\sim$ 2/3 of the total Fe yield over time
(more if the IMF slope or mass limits are changed to favor high
masses).  Thus, any element X with mass yield from VMS similar to Fe
should appear with the appropriate ratio [X/Fe] in the P inventory.

In the top panel of Figure 1 we compare the PISN yields (HW02) to the
EMP metal abundances from three sources (McWilliam et al. 1995;
Carretta et al. 2000; Cayrel et al. 2003).  We plot these samples
separately and mark the theoretical PISN yields for three progenitor mass
ranges with filled boxes.  Yields for unobserved elements are included
for completeness and marked with lighter colors. We see that the observed
abundances are consistent across the different studies for elements
they have in common. Clearly the high-mass PISN match well with Si,
Ca (as pointed out by Oh et al. 2001), and to some extent with Mn,
but they are a poor match to the other elements. They overpredict Cr,
underpredict V, Co, and produce little Zn.\footnote{HW02 point out that
PISN models may produce more Zn if neutrino-rich shocks are included in
future models. The potential for some future model to better match the
data does not presently constitute positive evidence of VMS.} The VMS
models are a poor fit to the data, especially for the Fe-peak elements Co
and Zn. The separation by mass range emphasizes that enrichment by one or
a few stars is a much poorer fit to the observed data than a mass-weighted
(IMF-averaged) yield. Some elements are matched best by the lower mass
range, while others favor the higher masses (which have Fe yields of up
to 40~\msun). Thus, the data appear to be inconsistent with a model,
such as the P inventory, in which these metal-poor stars formed from
gas that was enriched only once or a few times by VMS precursors. The
generally poor match of PISN yields leads us to conclude that VMS were
probably not the precursors of these EMPs.

The phenomenological model for $r$-process patterns by Fields, Truran, \&
Cowan (2002) does not make specific reference to VMS, but it is a good
example of another class of models that can test the VMS hypothesis.
This study explains the $\sim$ 2 dex scatter in \rfe\ by a stochastic
treatment of supernova enrichment that defines two types of SN with
identical Fe mass yield but very different \rfe\ (class A, high \rfe;
class B, low or zero \rfe).  Their artificial halo star sample reproduces
the observed scatter in [Eu/Fe] below \feh\ $= -2.5$ when A events
comprise $\lesssim$ 5\% of all SN.  In this model, A events produce
all the $r$-process elements, and B events make virtually all the Fe.
The assumed uniform Fe mass yield was adopted from SN 1987A (Woosley et
al. 1994), but it also matches quite well to the Fe yields of VMS PISN
from 140 -- 160~\msun.  The VMS PISN represent a good candidate for
the Fe-producing but $r$-process-poor B events, but in fact this model
illustrates the difficulty with any model that attempts to attribute the
EMP Fe budget to VMS. As shown above (and in Figure 1), the observed
abundances of Fe-peak elements in EMPs are not a good match to PISN.
Although the absolute Fe yields from VMS in the 140 -- 160~\msun\ range
serve well as B events, stars in this mass range produce negligible Zn.
The Fields et al. (2002) model is a generalization of the WQ00 P-inventory
model that accurately describes the observed scatter in \rfe\ at low
\feh. Both models thoroughly contradict the strong VMS hypothesis and
are inconsistent with the weak version. The Fields et al. (2002)
model is successful without including VMS, so the general class of models
it represents (including WQ00) does not provide positive evidence for VMS.

Some of the recent detections of Galactic halo stars with \feh\ $\leq
-3.0$ but enhanced [C,N,O/Fe] present another puzzle to nucleosynthesis
models. The lowest \feh\ star (HE0107-5240, \feh\ $ = -5.3$; Christlieb
et al. 2002) exhibits \cfe\ = 4.0 and \nfe\ = 2.3 (O was not measured),
so it may be more accurate to call this star ``iron poor'' than
``metal-poor''. Recent high-resolution analyses of CS 22949-037 (Norris
et al. 2002; Depagne et al. 2002) and CS 29498-043 (Aoki et al. 2003)
show similar patterns.  In their analysis of CS22949-037, Norris et al.
(2002) point out a good fit of \nfe\ = 2.6 to the rotating
zero-metallicity VMS models of Fryer et al. (2001), in which primary N
is produced by C and O mixing into the H-rich shell. However, they note
that this model overproduces N by a factor of 10 and that the star does
not exhibit the exaggerated odd-even nuclei effect of PISN.  In a
subsequent analysis of \ofe\ in this same star, Depagne et al. (2002)
argue for a close match to an unpublished model of Heger \& Woosley
with $M = 35$~\msun.  This model (Z35Z) matches the observed abundances
of CS22949-037 when modified to include NLTE effects for Al and Na and
enhanced N production by C and O mixing into the H-rich shell.
Furthermore, this model produces Zn in the correct ratio to Fe, while
no PISN models can do so.

In a study of HE0107-5240, Limongi et al. (2003) argue that this star
cannot be fitted with the yields of a single supernova because the
Fe-peak elements favor a deep mass cut (such that relatively more
material is ejected), while the light elements favor a shallow one (to
produce high [C/Fe]). Instead, they obtain a good fit for two SN
sources: an earlier 15~\msun\ SN with a deep mass cut between the
ejected material and the remnant that releases the heavier elements,
and a later 35~\msun\ SN with such a shallow mass cut that the
supernova nearly fails to eject any metals at all. Although it requires
two fine-tuned SN models to do so, their model matches HE0107-5240
without recourse to VMS. Umeda \& Nomoto (2003) also find that the
abundance pattern in this star fits nucleosynthesis from metal-free
stars with $M = 20 - 130$ \msun.

Because $r$-process enrichment is common in EMP stars, and because PISN
experience no $r$-process (HW02), we can rule out the strong VMS
hypothesis; clearly  8 -- 40~\msun\ stars must have existed in the
generation that enriched the EMP stars.  Furthermore, because the
existing models with plausibly constructed IMFs in the ``weak'' VMS
formulation must attribute all the early Fe to VMS, they also fail to
match the observed abundance patterns of EMP stars.  Thus, we find that
the nucleosynthesis constraints do not provide concrete, positive evidence
for VMS that cannot be explained by other means.  Although ambiguities
remain in the interpretation of the CNO-rich EMPs, they clearly do not
{\em require} VMS and/or PISN. Unless additional input physics in PISN
models change the odd-even and Fe-peak characteristics of their yields,
it appears the EMP stars do not support the VMS hypothesis.

\subsection{Hypernovae and Nucleosynthesis}

Umeda \& Nomoto (2002, 2003, 2004), and Nomoto et al. (2003) have
developed an alternative explanation for EMP metal abundances. They argue
that energetic supernovae or ``hypernovae'' (HN; $E_{51} = 10 - 100$,
where $E_{51} = E/10^{51}$ erg) that appear to be connected with some
low-$z$ gamma-ray bursts could be relevant for early nucleosynthesis if
they exist at low metallicity.  Empirical estimates based on light curves
(Hamuy 2003) suggest that these events arise from supernovae with $E_{51}
= 10 - 30$ and progenitors of $M \gtrsim 15$~\msun. Since they have been
found at low redshift, these objects are somewhat less speculative than
VMS and PISN, which currently have only a theoretical motivation.

Umeda \& Nomoto (2004) argue that the trends in the Fe-peak elements in
EMP stars reflect nucleosynthesis by HN from zero-metallicity
progenitors of $M < 140$ \msun.  In their models, the joint effects of
a deeper mass cut between remnant and ejecta, mixing in the ejected
material, and the fallback of some of the metals onto the remnant
enhance the products of complete Si-burning (Zn, Co) relative to the
products of incomplete Si-burning (Mn, Cr). A higher explosion energy
also increases the dilution mass (the mass of pristine gas into which the
ejected metals are mixed).  Thus, second-generation stars born of their
products will possess lower \feh\ than stars enriched by lower-energy
SN, with higher [(Zn, Co)/Fe], and lower [(Cr, Mn)] as \feh\ decreases.
This detail in the models may need to change in light of the tighter
scatter and smoother trends in the EMP observations of Cayrel et
al.~(2003). However, in general, HN appear to provide a superior match to
the EMP data.  Furthermore, Umeda \& Nomoto (2004) argue that CNO-rich
EMP stars can be explained with variations in explosion energy and the
degrees of mixing and fallback in the ejected material.

For comparison to the PISN yields, the lower panel of Figure 1 shows
the HN models compared to the three EMP surveys.  We display the HN
yields for the Fe-peak elements Cr, Mn, Co, and Zn resulting from
explosions with $M = 13 - 50$~\msun\ and $E_{51} = 1 - 100$ (UN04).
These elements are insensitive to the location of the mass cut in the
supernova explosion, but they can vary with the electron mass fraction
($Y_e$) in the ejected material, the effects of which we include in the
plotted range to show the improved fit (e.g., Figure 4 of Umeda \&
Nomoto 2004). The large explosion energies of hypernovae are needed to
reproduce the observed Zn. For all other elements, we plot the yields
of Umeda \& Nomoto (2002), with $M = 15 - 20$~\msun, $E_{51} = 1 - 5$,
and including mixing and fallback of the ejected material.  While the
hypernova phenomenon is gaining support from gamma-ray burst
observations and theory, the HN explosion energy is still a free
parameter in the Umeda \& Nomoto (2004) models, and it is somewhat
degenerate with stellar mass as a determining factor of the metal
yields.  There is also uncertainty associated with extrapolation from
the low-$z$, roughly solar-metallicity conditions to $Z \leq 10^{-3}
Z_{\odot}$. Although the apparent match of HN to the EMP metal
abundances is encouraging, they cannot yet be conclusively identified
as the products of the first stars. Nevertheless, the possible
existence of HN with $M = 10 - 50$ \msun\ in the first generation is an
important part of constructing a first-stars IMF without VMS.

\subsection{Discussion of Nucleosynthesis Results}

We have reviewed the available evidence from stellar nucleosynthesis
studies that bear on the mass function of the first stars. Oh et
al. (2001) proposed that the correspondence between the HW02 PISN yields
and the WQ00 prompt inventory implied that VMS were the sources of Fe,
Si, and Ca at \feh\ $< -3$. They used this Fe abundance to estimate the
global ionizing photon budget available to reionize the universe. These
conclusions are not supported by the broader sets of elemental abundances.
In contrast, we draw the following conclusions from our examination
of the nucleosynthetic evidence: \begin{itemize} \item[1.] The common
presence of \rp\ elements in halo EMPs (McWilliam et al.
  1995; Burris et al. 2000) suggests progenitors of 8 -- 40~\msun, where
  the $r$-process elements are thought to form (Truran et al.  2002).
  This requirement is incompatible with the strong VMS hypothesis and
  emphasizes the importance of the lower mass limit of the IMF.
\item[2.] The relative abundances of Fe-peak elements in EMPs
  (especially Zn; Cayrel et al. 2003) imply Type II supernovae from
  massive stars, or energetic hypernovae with deep mass cuts and mixing
  in the ejected material (Umeda \& Nomoto 2004). These abundances are not
  well-fitted by the yields of pair-instability supernovae from VMS. This
  is not a refutation of the weak VMS hypothesis, because it does not
  constitute evidence that VMS were not present at some level. It does
  imply that VMS are not responsible for the bulk of the Fe at \feh\ $<
  -3$ (as suggested by WQ00 and Oh et al.~2001).
\end{itemize}

Much depends on the true nature of the Galactic halo EMP stars.  If
they do not, in fact, trace the enrichment of the first stellar
generation, then we will be left with few, if any, such tracers and
none that indicate VMS.  For example, damped Ly$\alpha$ systems (DLAs)
in QSO spectra are often considered to be candidates for studies of
primordial nucleosynthesis, because they have low metallicity and, at
high $z$, they may reflect early chemical enrichment. However, the
existing literature on DLAs (Prochaska \& Wolfe 2002 and references
therein) shows that they generally have \feh\ $ > -2$, while the
Galactic halo EMPs range down to $-4$.  Thus, they may not reflect the
same primordial metal enrichment thought to be seen in the EMPs with
\feh\ $\leq  -3$. Furthermore, the selective depletion of Fe and other
refractory elements onto dust grains in DLAs can change the Fe-peak
abundance ratios in the gas and thus complicate comparisons to
theoretical yields and other observations. Finally, direct ratios
between Zn, Cr, and Mn in DLAs match the EMP stars that also have \feh\
$> -2$, above the metallicity where the qualitatively distinct behavior
of EMPs appears. While Galactic stars and DLAs in this higher
metallicity regime may derive their metals from similar sources, and
may in fact have unusual features, DLAs by themselves do not indicate
that they have been enriched by PISN.  The presence of VMS is also not
required by either the observed metal ratios in QSO broad emission line
regions (Venkatesan, Schneider \& Ferrara 2004) or by [Si/C] ratios 
in high-redshift Lyman limit systems (Levshakov et al. 2003).

The recent studies (see \S~2) of reionization by VMS have expanded on
the reionization thesis offered by Oh et al. (2001), but they have
relied for the most part on that initial study to ground their use of
the VMS hypothesis in terms of the nucleosynthetic evidence. Although
it has been evaluated by observers in their presentation of the data,
the independent evidence for the VMS hypothesis has not been widely
examined in the theoretical literature. Previous studies have in many
cases met the {\it WMAP} and Gunn-Peterson constraints on the H~I
reionization history of the IGM, but they have done so by assuming
extreme astrophysical parameters or by assuming the VMS hypothesis,
which we find to have no firm external justification. We return to the
issue of reionization by the first stars in \S~5.

\section{The IMF of the First Stars}

\subsection{Constructing an IMF}

With the nucleosynthetic results in mind, we can now construct a
first-stars IMF that is consistent with the observations.  To the
observational constraints discussed in \S~3.4 we add the following
theoretical motivations for constructing an IMF that is deficient in
low-mass stars but not skewed to the VMS mass range:
\begin{itemize}
\item[3.] The general astrophysical arguments (cooling, etc.) used to
   motivate a ``top-heavy'' IMF still apply. Apparently forming massive
   stars at $Z = 0$ is a fundamentally different process than it is today.
   While Tan \& McKee (2003) point out that formation feedback may limit
   the mass of metal-free stars to $\sim$ 30~\msun, they also suggest
   that there is no known mechanism for halting the accretion below this
   range ($\sim 1 - 10$~\msun) or fragmenting into smaller clumps. On general
   grounds, this justifies increasing the lower limit of the IMF.
\item[4.] As shown in Figure 2,
   the ionizing efficiency of metal-free stars increases rapidly from
   1 -- 50~\msun, with a plateau at $100 - 120$ \msun.  Thus, we can
   construct an IMF that achieves an ionizing efficiency similar to even a
   ``strong'' VMS model without recourse to stars with $M \geq 140$ \msun.
\end{itemize}

We propose a first-stars IMF that embodies the observational constraints
(1, 2) and the theoretical arguments (3, 4). It is consistent with the
available nucleosynthetic evidence, and it maximizes the ionizing photon
efficiency to meet existing constraints on the reionization without
violating those on IGM metal enrichment below $z \la 6$ (Venkatesan \&
Truran 2003).  In Table 1 we present six new IMF cases that meet these
criteria.  There are four cases with Salpeter slope and varying upper
and lower mass limits (B -- D, G), and two cases with varying slope (E --
F) to capture the effects of proportionally more massive star formation
in addition to variable mass limits. We also include a pure-VMS IMF (A)
for comparison. Although it is tempting to change the functional form
of the IMF from a power law to a more complicated function that favors
high-mass stars, we avoid this for two reasons. First, the formation
models have not given us specific reason to do so; they suggest that the
mass limits could change but have not yet determined the shape of the
distribution. Second, a more general function departing from a power law
would introduce another parameter, and the IMF slope is already poorly
constrained.  Even if one knows the IMF slope or mass limits, but not
both, the inherent degeneracies in the problem may permanently preclude
well-justified constraints on the other.  A major part of the IMF ($M =
50 - 140$~\msun) leaves little or no nucleosynthetic trace, and so the
relative contributions of these stars to the ionizing radiation cannot
be easily constrained.  Thus, there remain significant uncertainties
and ambiguities in constraining the first-stars IMF.

Figure~\ref{phot} shows why our proposed IMF cases can achieve
reionization results similar to VMS. Here we show the
lifetime-integrated ionizing photon production per baryon (``ionizing
efficiency'') in metal-free stars of 5 -- 500~\msun. Up to $M =
100$~\msun, the models are from Tumlinson, Shull, \& Venkatesan (2003);
for $M > 100$~\msun\ they are from unpublished models based on the same
evolution and stellar atmosphere code, checked for consistency against
the results of Schaerer (2002). The notable features of this plot are
the sharp rise to a peak at $M \simeq 120$~\msun, followed by a slow
decline at higher mass. This effect is a consequence of the stellar
lifetimes decreasing more rapidly than the main-sequence ionizing
photon production increases.  This curve shows that an IMF that
excludes low mass stars and their poor ionizing efficiency can closely
approximate the total ionizing photon production of an IMF composed
purely of VMS.

Recently, Venkatesan \& Truran (2003) showed that VMS may not be
responsible for both reionization and IGM metal enrichment, since they
produce only about 0.35 ionizing photons per baryon before they cease
forming due to metal pollution (reionization at $z \sim 6$ is thought
to require $\sim 10$ ionizing photons per baryon in the IGM, and more
are required at higher redshifts).  This contradiction would be
resolved if the first stars were all VMS with $M > 260$~\msun\ and
reionization was accomplished by stars that emit ionizing radiation but
release no metals. However, this way is probably closed because no
mechanism has been proposed for forming 300~\msun\ stars without also
forming 200~\msun\ stars in roughly equal or greater proportion.
Furthermore, a primordial IMF composed exclusively of $M > 260$~\msun\
stars would never lead to the observed present-day star formation, as
metals would not be released to trigger the transition from the first
stars to a local IMF (the enrichment paradox; Schneider et al. 2003).
The only way out of this paradox is to form stars below $M =
260$~\msun, where metals would be released by PISN and influence later
star formation. Although this argument could be modified by future
results from formation models, this probably means that excluding 140
-- 260~\msun\ also rules out more massive objects.

\subsection{The Duration of Metal-Free Star Formation}

A key parameter in modeling the radiative and chemical input from the
first stars is the duration of the ``metal-free" phase of star
formation. Given the complications of an unknown IMF, interstellar
mixing, and halo merging history, this question is difficult to pose
quantitatively. Here we make simplifying assumptions to  estimate the
enrichment time of small halos at $z = 10-20$ to estimate the time
during which metal-free gas is available to form stars.

As noted previously (Tumlinson, Shull, \& Venkatesan 2003; VTS03), the
duration of the metal-free phase determines, with the IMF, the total
output of ionizing photons from the first massive stars.  As
they die, these stars eject heavy elements from supernovae and enrich
the surrounding gas within their own halo. These metals may also be
transported across intergalactic space to enrich neighboring halos.
Although these enrichment and transport processes are intrinsically
complicated, we can make a few relevant calculations that constrain the
duration of this metal-free epoch.  Our calculations are similar in
spirit to those of Scannapieco, Schneider, \& Ferrara (2003), but we
make different assumptions.

First, we compute the {\it global} rate of metal enrichment per unit
volume, based on an average star formation rate density (SFR),
hereafter quoted in standard units of comoving $M_{\odot}~{\rm
yr}^{-1}~{\rm Mpc}^{-3}$. By finding high-redshift galaxies through B,
V, and $i$-band dropouts, Steidel et al.\ (1999) and Giavalisco et al.
(2004) estimated a SFR $\approx 0.1$ between $3 < z < 6$.  From the
required rate of ionizing emission to reionize the IGM by $z = 6$,
Miralda-Escud\'e (2003) showed that the ionizing (comoving) emissivity
cannot decline by more than a factor of 1.5 from $z = 4$ up to $z = 9$.
Recent numerical simulations of galaxy formation (Gnedin 2000; Ricotti,
Gnedin, \& Shull 2002a,b, hereafter denoted RGS) allow us to
extrapolate to the earliest epochs of galaxy formation ($z \approx
20-30$) when star formation begins through H~I and H$_2$ cooling of
protogalactic halos, with virial temperatures $T_{\rm vir} = 10^3$~K to
$10^{4}$~K. RGS found that the average SFR increases monotonically,
from $10^{-3}$ to $10^{-1}$  between $z = 20$ and 10, according to the
approximate formula:
\begin{equation}
   {\rm SFR} \simeq (0.003~M_{\odot}~{\rm yr}^{-1}~{\rm Mpc}^{-3})
       10^{(20-z)/5} \; .
\end{equation}
These simulations agree fairly well with the extended Press-Schechter
estimates of Scannapieco et al. (2003).  To estimate the corresponding
metal production, we adopt a fiducial SFR = $0.01~M_{\odot}~{\rm
yr}^{-1}~{\rm Mpc}^{-3}$, with standard metal yield, $y_m = 0.00663$
for metal-free stars of $M = 1 - 100$ \msun.  After a time $t =
(10^8~{\rm yr}) t_8$, the average IGM metallicity is:
\begin{equation}
   \frac {Z_{\rm IGM}} {Z_{\odot}} = \frac {({\rm SFR}) \, y_m \, t}
         {\Omega_b \rho_{\rm cr} (0.02)}
     = (5.4 \times 10^{-4}) \left( \frac {{\rm SFR}} {0.01} \right) t_8 \; .
\end{equation}
The elapsed time between $z = 20$ and $z = 10$ is $2.9 \times 10^8$ yr
for $H_0 = 70 h_{70}$ km~s$^{-1}$, $\Omega_m = 0.3$, and
$\Omega_{\Lambda} = 0.7$.  The above estimate shows that early star
formation can raise the {\it average} IGM metallicity to values near
the suggested critical threshold, $Z_{\rm cr} \approx 10^{-3.5}
Z_{\odot}$ at which metal cooling may produce a transition to a
different mode of star formation involving fragmentation into low-mass
clumps (Omukai 2000; Bromm \& Loeb 2003; Schneider et al. 2002).

Complicating this estimate is the likelihood that primordial metal
enrichment proceeded inhomogeneously (Scannapieco et al. 2002, 2003).
Here, we consider a halo with total mass $M_h = (10^6~M_{\odot}) M_6$,
in which $1.66 \times 10^5 M_6$ $M_{\odot}$ of baryons are distributed in
$N$ clumps spread over a halo of virial radius $R_{\rm vir}$ and virial
temperature $T_{\rm vir}$:
\begin{eqnarray}
  R_{\rm vir} &=& \left[ \frac {3M_h} {72 \pi^3 \rho_m} \right]^{1/3}
     = (165~{\rm pc}) M_6^{1/3} \left[ \frac {20}{(1+z)} \right]  \\
      \nonumber    \\
  T_{\rm vir} &=& \left[ \frac {G M_h m_H} {3 k_B R_{\rm vir}} \right]
      = (1060~{\rm K}) M_6^{2/3} \left[ \frac {(1+z)}{20} \right] \; .
\end{eqnarray}
We assume cosmological densities, $\rho_m = (1.88 \times
10^{-29}h^2~{\rm g~cm}^{-3}) \Omega_m (1+z)^3$, with $\Omega_m h^2 =
0.135$ and $\Omega_b h^2 = 0.0224$ (for matter and baryons,
respectively; Spergel et al. 2003) and a virialized density ratio
$\rho(R_{\rm vir})/\rho_m = 18 \pi^2$.  We set halo radius $R_h =
R_{\rm vir}$ and divide the matter into $N$ clumps, with contrast ratio
$\rho_{\rm cl}/ \rho_h \gg 1$, where $\rho_h = (3M_h/4 \pi R_h^3)$. We
next assume that 5\% of the halo's baryons undergo a burst of star
formation at the center of the halo.  As these stars eject their
metals, a small fraction, $f_{\rm cap}$, of these metals are captured
by the nearest-neighbor clumps, while most are blown out into the IGM.
The capture fraction may be estimated from the solid angle of the
neighboring clump, of radius $r$, as seen from the halo center, $f_{\rm
cap} \approx (\pi r^2 / d_{\rm cl}^2)/(4 \pi)$.  The mean distance
between clumps within the halo is $d_{\rm cl} \approx (3N/4 \pi
R_h)^{-1/3}$. We also have $N r^3 \rho_{\rm cl} = R_h^3 \rho_h$.
After some algebra, we find
\begin{eqnarray}
  f_{\rm cap} &=&  \left[ (1/4) (3/4 \pi)^{2/3} \right]
    (\rho_h / \rho_{\rm cl})^{2/3}
   \nonumber \\
    & \approx &   (0.096)(\rho_h/\rho_{\rm cl})^{2/3} \; .
\end{eqnarray}
We arrive at the simple result that the fraction of metals
captured by the nearest-neighbor clump depends only on the
density contrast fraction between the clump and halo.

For example, for the $10^6~M_{\odot}$ halo, with total baryonic mass
$M_b = 1.66 \times 10^5~M_{\odot}$, we assume $N = 30$ clumps, each
with $5530~M_{\odot}$ in baryons and a clump-interclump contrast ratio,
$\rho_{\rm cl}/\rho_h \approx 100$. If 5\% of the halo's baryons
($8300~M_{\odot}$) undergo a burst of star formation at the halo
center, they will produce approximately 83 SNe and $18.3~M_{\odot}$ of
heavy elements ($y_m \approx 0.00663$).  The nearest-neighbor clump
will capture a fraction $f_{\rm cap} = 4.5 \times 10^{-3}$, for an
``enriched" metallicity of $(18.3)(4.5 \times 10^{-3})/(5530) \approx
1.5 \times 10^{-5}$ or $7.4 \times 10^{-4}~Z_{\odot}$.  Thus, the
sub-clumps within the halo will be enriched to $\sim 10^{-3}$ solar
abundances on a timescale of a few Myr, the crossing time of the SNe
ejecta across the halo.

A longer, and perhaps more relevant, timescale is that required to
transport the heavy elements between halos, across the IGM. From recent
simulations (Gnedin 2000; RGS), the co-moving density of $10^6~M_{\odot}$
star-forming halos at $z = 15-20$ is approximately $4 \pm 2$ Mpc$^{-3}$,
corresponding to a mean (physical) distance between halos of $d_{\rm halo}
\approx (30~{\rm kpc})[(20/(1+z)]$.  Thus, the ejected heavy elements, if
they are transported at mean velocity $V_f \approx 100~{\rm
km~s}^{-1}$, would not reach neighboring halos for over $10^8$ yrs.

These neighboring halos may recede with the local Hubble flow, $H(z) =
(3.43~{\rm km~s}^{-1}~{\rm kpc}^{-1}) h_{70}$ at $z = 20$. The Hubble
expansion velocity between these halos could be as large as
\begin{eqnarray}
   V_{\rm exp} & \approx & H_0 (1+z)^{3/2} \Omega_m^{1/2} d_{\rm halo}
      \nonumber \\
     & \approx & (100~{\rm km~s}^{-1}) \left[ \frac {1+z}{20}
        \right]^{1/2} \; .
\end{eqnarray}
Alternatively, this region of space could have turned around from the
expansion, depending on the extent and spatial correlations of the mass
perturbations. Transporting the ejected metals across 30 kpc at $z =
20$ is not easy, given the amount of intervening intergalactic matter.
A sphere of 1 kpc (physical distance) surrounding a halo at $z = 20$
contains $2 \times 10^5~M_{\odot}$ of baryonic mass, at the mean
hydrogen number density,
\begin{equation}
  n_H = (1.52 \times 10^{-3}~{\rm cm}^{-3})
                    \left[ \frac {1+z}{20} \right]^3 \; .
\end{equation}
Any heavy elements ejected in SNe explosions or outflows are accompanied
by significant energy deposition. They rapidly sweep up many times
their own mass, form cosmological blast waves, and eventually develop
radiative shells (Ostriker \& Cowie 1981; Shull \& Silk 1981) which slow
down as they expand.

Consider the 5\% burst of star formation from the previous example.
The resulting 83 SNe would deposit an energy $E_{\rm SN} \approx
(10^{54}~{\rm erg})E_{54}$ into the IGM, if we assume an
energy of $10^{52}$~erg per SN or HN (Umeda \& Nomoto 2002). An estimate of the
blast wave extent comes from the non-cosmological Sedov-Taylor formula,
\begin{eqnarray}
  R_s &=& \left[ \frac {2.02 E_{\rm SN} t^2} {\rho_{\rm IGM}}
        \right]^{1/5}
    \nonumber \\
      & \approx& (7.35~{\rm kpc}) \left[ E_{54} t_8^2 /\delta \right]^{1/5}
        \left[ \frac {1+z}{20} \right]^{-3/5}   \; .
\end{eqnarray}
Here, $\rho_{\rm IGM} = 1.32 n_H m_H$ is the IGM mass density, $\delta$
is the overdensity relative to the mean, and $t = (10^8~{\rm yr})t_8$
is the time after the injection of SN energy.  In reality, one must
take cosmological expansion and IGM inhomogeneity into account, as well
as the gradual injection of energy into the ``superbubble" produced by
multiple SNe over a total time of $\sim 5 \times 10^7$~yr.

From the above approximate formulae, we see that the expanding blast
wave could take over $10^8$ yr to contaminate neighboring halos. We
conclude that the halo-to-halo enrichment timescale is likely to be of
order $10^8$ yr. These simple calculations motivate our choices of
$10^7$ and $10^8$ yr as the duration of the first stars epoch for
primordial star formation. These timescales are in accord with those
calculated from numerical simulations of halo self-enrichment by Wada
\& Venkatesan (2003) and of interhalo enrichment by Bromm, Yoshida, \&
Hernquist (2003). The effects of these two timescales on the
reionization models are discussed in \S~5.

In fact, our proposed first-stars IMF has another virtue along these
lines.  If most of the mass in the early IMF is formed into stars in
the 50 -- 140~\msun\ mass range, the unique effects of a
zero-metallicity IMF could persist longer than they would if the IMF
favored VMS. To see why, consider a case where all mass goes into
forming stars with either $M = 50 - 140$~\msun\ or $M > 260$~\msun.
Since these stars collapse directly to black holes without releasing
metals, the metal-free mode of star formation would be
self-perpetuating. If this extreme IMF is modified to incorporate
increasing but still relatively small numbers of SN II or HN
progenitors, then metal enrichment will gradually achieve the global
value of {\bf $Z = 10^{-3.5} Z_\odot$} thought to curtail the unique
low-metallicity star formation mechanisms. In a rough sense, this
timescale is inversely proportional to the fraction of matter formed
into stars that explode and release metals relative to those that do
not.  Thus, a significant advantage of the $M = 30-140$ \msun\ IMF is
that the duration of metal-free star formation at early epochs could be
longer, owing to the increased stellar lifetimes in this IMF relative
to VMS and to the lower total metal yields.

\section{The First Stars and Reionization}

We have derived a first-stars IMF that is consistent with the available
nucleosynthetic evidence, and we have estimated the time ($\sim 10^7 -
10^8$ yr) over which metal-free stars can form in small primordial
halos. We must now evaluate our proposed IMF in terms of reionization
models to compare their effect on the IGM to that of VMS. We hold fixed
all the model assumptions except the IMF and the metal-enrichment
timescale to determine the extent to which the IMF matters to the
outcome.

We adopt the semianalytic stellar reionization model described in VTS03
for a concordance $\Lambda$CDM cosmology. The growth of ionized regions is
tracked by a Press-Schechter formalism in combination with numerical
solutions for the growth of individual ionization fronts.  For all cases
here, the cosmological and astrophysical parameters are the same as in
VTS03; we do this in order to consistently compare the effects of a VMS
IMF relative to those proposed here (there is one exception, mentioned
below).  Specifically, our models are described by the parameter set
[$\sigma_8$, $n$, $h$, $\Omega_b$, $\Omega_\Lambda$, $\Omega_{\rm m}$,
$c_L$, $f_\star$, $f^{\rm H}_{\rm esc}$, $f^{\rm He}_{\rm esc}$] = [0.9,
1.0, 0.7, 0.04, 0.7, 0.3, 30, 0.05, 0.05, 0.025], where $\sigma_8$
and $n$ are the usual normalization and index of the scalar matter
power spectrum, $c_L$ is the space-averaged clumping factor of H~II or
He~III (assumed to be the same here), $h$ is the Hubble constant in units
of 100 km s$^{-1}$ Mpc$^{-1}$, and $\Omega_b$, $\Omega_\Lambda$, and
$\Omega_{\rm m}$ represent the cosmological density parameters of baryons,
the cosmological constant, and matter respectively. The astrophysical
parameters are $f_\star$, the fraction of baryons forming stars in each
halo, and $f^{\rm H}_{\rm esc}$ and $f^{\rm He}_{\rm esc}$, the escape
fractions of H~I and He~II ionizing photons from halos, respectively.

We allow star formation in all halos of virial temperature $\ga$ 10$^3$~K
(rather than 10$^4$~K as in VTS03), so as to have sufficient baryons
in a $\sim 10^6 M_\odot$ dark matter halo at $z = 20$ for our adopted star
formation efficiencies to form the most massive star ($\sim 10^3 M_\odot$)
in our IMF cases. Metal-free stars are assumed to form starting from $z
\sim 20$ and last in each halo for either of two approximate timescales
derived earlier for intrahalo and halo-halo enrichment, $10^7$ yr and
$10^8$ yr. Subsequent to this, the ionizing spectrum is switched to a
representative example of Pop II stars with metallicity $Z = 0.001$
(Leitherer et al. 1999). Reionization is defined as the overlap of
individual ionized regions of H~II or He~III as appropriate, when their
volume filling factors equal unity (see VTS03).

Relative to the results of VTS03, reionization in our new models occurs
earlier, owing to the lower halo mass threshold and the fixed
metal-free star forming phase in each halo before switching to a Pop II
ionizing spectrum. VTS03 assumed $T_{\rm vir} \geq 10^4$~K and shut off
metal-free star formation at fixed cosmological ages.  Thus, in VTS03,
the majority of small halos, in which most of the baryons reside until
$z \la$ 9 when they merge into larger systems, did not experience any
metal-free star formation, even if they had not undergone prior star
formation or experienced pollution from neighboring halos. Our
treatment here ensures that each halo experiences at least one episode
of metal-free star formation prior to cosmological times of $\sim 1$
Gyr. The assumption of $T_{\rm vir} \geq 10^3$~K is somewhat more
optimistic from the point of view of early reionization, and relies on the
assumption that halos with $T_{\rm vir} \leq 10^4$~K can effectively cool
with H$_2$ (RGS). Other models (e.g., Haiman \& Holder 2003) have even
more optimistically assumed star formation in halos down to $T_{\rm vir}
\geq 10^2$~K; these are not massive enough to form the most massive star
in our IMF cases (1000 \msun).  The assumption that each new halo forms
metal-free stars breaks down at $z \simeq 9$, when the earliest small
halos, presumably metal-enriched from their first star formation, begin to
coalesce into larger objects, which are presumably never metal-free. Our
models reionize early enough to avoid this complication (see below). 

In Table 1, we display the various IMFs considered in this work. We
include newly calculated cases that are motivated by the
nucleosynthetic data (\S~4.1) and cases that represent a top-heavy IMF,
either in the form of a higher stellar mass range or of a flatter IMF
slope, $\alpha$. Also shown are the ionizing photon rates, $Q_i$, for
H~I and He~II from synthetic stellar clusters at a starburst age of 1
Myr in units of photons s$^{-1} M_{\odot}^{-1}$. We also show the H~I
and He~II reionization redshifts, $z_r$, for each case and the total
associated $\tau_e$ (including the $\sim 10 - 15$\% contribution from
He~II), assuming that the post-recombination IGM ionization fraction is
about $10^{-4}$ over $z \sim$ 20 -- 1000.  The equivalent ionizing rates
for Pop II are $Q_{\rm HI} \sim 7.8 \times 10^{46}$ s$^{-1}$ \msun$^{-1}$
and $Q_{\rm HeII} \sim 1.3 \times 10^{40}$ s$^{-1}$ \msun$^{-1}$
after the transition from Pop III to Pop II\footnote{For
comparison, the reionization models presented by VTS03 produced $z_{r,H}
\simeq 7$ assuming a Salpeter IMF with 1 - 100 \msun, halos with $
T_{\rm vir} > 10^4$ K, and a Pop III to Pop II transition time of
10$^8$ yr.}. We see that an IMF derived from the EMP nucleosynthetic
constraints and lacking in low-mass stars can generate sufficiently
early reionization and values of $\tau_e \geq 0.10$. This is consistent
with current data on reionization without resorting to a top-heavy
IMF; high ionizing efficiency need not arise from VMS only. We also
note that the uncertainty on $\tau _e$ from {\it WMAP} arises from a
broad probability distribution, and that it is degenerate with $n_s$.
Recently, $\tau _e = 0.12^{+0.08}_{-0.06}$ has been determined for {\it
WMAP} plus the three-dimensional matter power spectrum from SDSS (Tegmark
et al. 2003). Our models reproduce the {\it WMAP} result at the $1 \sigma$
confidence level and match the Tegmark et al. (2003) value without resort
to VMS. In fact, the close agreement achieved by our IMF suggests that the
astrophysical parameters (star formation efficiency, escape fraction) are
now the limiting factors in achieving good constraints on reionization.
We also note that the models from VTS03 produced $\tau_{\rm e}
\sim 0.08$ with a more restrictive halo virial temperature cutoff ($10^4$
K) and a Salpeter IMF. The difference between this $\tau_{\rm e}$ and
model B shows that the more optimistic halo prescription of the present
study contributes only a modest amount. 

An important caveat for the results here is that the IGM, once reionized
completely in H or He, stays ionized.  For H~I, if reionization occurs
once in the model at some redshift, e.g., $z = 15$, and if metal-free
stars are allowed to keep forming subsequently for $10^7$--$10^8$ yr
in each new halo, the model IGM does not recombine and stays ionized,
even though recombinations are explicitly allowed. This result is
probably valid for H~I in the underdense voids, which represent most of
the volume in the IGM. In the case of H~I in dense regions and He~II in
general, this behavior may be suspect, owing to high recombination rates
at typical IGM densities at $z \ga$ 5 (Shull et al. 2004). However,
even if He~III partially recombined in those of our cases in which
He~II reionization occurs early, the contribution to $\tau_e$ would
be impacted at less than the 10\% level. Therefore, our numbers for
He~II should be considered only for the purposes of comparison between
various IMFs. This is especially true at $z \lesssim 6$ where we have not
included the ionizing radiation from QSOs.  The simple reionization
history in our models also does not reproduce the known H I Gunn-Peterson
opacity at $z \sim 6$ or the observed IGM temperatures at $z = 2 -
4$. These constraints would require additional input into the models,
such as the contribution of quasars to the ionizing photon budget,
or a more complicated reionization history than we explore here.
Models that attempt to match all available constraints exist in the
literature (see \S~2). The additional complications and new parameters
would, if they were included in our model, needlessly distract from the
goal of directly comparing the different IMF cases with the same minimal
set of cosmological and astrophysical parameters. We also note that the
$z \simeq 6$ Gunn-Peterson constraint can be met by a modest amount of
recombination at $z = 7 - 9$, perhaps coincident with the expected Pop III
to Pop II transition, with only a modest effect on $\tau_{\rm e}$. This
is possible because even a trace neutral fraction is sufficient to provide
$\tau_{\rm GP} \sim 1$ with negligible impact on the ionized fraction
that scatters photons from the CMB. Because these external constraints
can be met with minor adjustments to our model, and because they are
incidental to the IMF comparisons, we do not include them here. 

\section{Conclusions and Discussion}

We close by drawing some important conclusions from the foregoing results
and discussion. These are:
\begin{enumerate}
\item The VMS hypothesis should be distinguished into two
      versions: the ``strong'' version (``the first stars were
      exclusively VMS'') and the ``weak'' version (``the first stars
      included VMS and stars with $M < 140$~\msun''). This distinction
      is necessary because the radiative and nucleosynthesis properties
      of metal-free stars vary widely across the mass function. We have
      demonstrated that these two versions have different testable
      predictions and can be independently falsified.
\item The existing samples of metal abundances in EMP
      Galactic halo stars, which are commonly thought to trace the
      enrichment products of the first generation of stars, are better
      matched to the products of Type II SN and hypernovae like those
      seen at low redshift. Specifically, the Fe-peak and $r$-process
      abundances in EMPs are not well-matched by VMS.
\item Because they are efficient sources of ionizing photons, VMS have
      been proposed as the ionizing sources for the first stage of
      the IGM reionization and the only means of generating large
      electron-scattering optical depths seen in CMB maps. By excluding the
      low-mass end of the IMF, as suggested by models of primordial
      protostars, we have constructed IMF cases that are just as
      efficient as VMS at converting baryons to ionizing photons.
      Thus constraining the lower mass limit of the IMF is perhaps more
      important for reionization than a constraint on the upper limit.
\item Direct comparisons of our test
      IMFs in semi-analytic reionization models show that without VMS
      these models can reproduce $\tau _e$ in the observed range, and
      similar to VMS models with the same astrophysical parameters. We
      can simultaneously reproduce the nucleosynthesis and reionization
      constraints with a single IMF. We confirm earlier studies (VTS03, 
      Wyithe \& Loeb 2003) which found that the lower end of the 
      {\it WMAP} range of $\tau_{\rm e}$ can be reproduced without recourse 
      to a nonstandard IMF.  
\item With simple metal-transport
      calculations, we derive a timescale for metal-free star formation of
      $\sim 10^7 - 10^8$ yr. By effectively decoupling the earliest stages
      of ionization and metal-enrichment, our proposed IMF can allow
      the period of metal-free star formation to persist longer than
      an IMF composed of VMS, and perhaps long enough to fully reionize
      the IGM before self-enrichment of halos permanently curtails
      metal-free star formation.
\end{enumerate}

We have shown that the nucleosynthesis constraints exclude the strong
VMS hypothesis and have difficulty accommodating the weak VMS
hypothesis. We have also shown that an IMF without VMS can reproduce
$\tau _e \sim 0.08$ if it includes stars with $M \lesssim 10$ \msun\
and $\tau _e \sim 0.10 - 0.14$ if it does not. Rather than provide
strong evidence for the VMS hypothesis, the existing $\tau _e$ values
instead argue against low-mass stars but provide little constraint on
massive stars in the IMF. In light of the metal-free star formation
models, both the high- and low-mass ends of the stellar mass function
are poorly constrained. It is only when the nucleosynthetic information
is added that we can say anything about the IMF.

With these uncertainties in mind, we consider how future developments
might bear on the first-stars IMF. First, a tighter range of $\tau _e$
is necessary to constrain the ionizing efficiency of the first stars,
and, indirectly, the IMF. Such an improvement might come with
additional {\it WMAP} data, including polarization correlations.
However, the greatest need is for consistent results from the various
combinations of the CMB and additional data (i.e., matter power
spectrum, Ly$\alpha$ forest) that can constrain $\tau_e$. Uncertainty
of $\sigma_{\tau} \lesssim 0.02$ on $\tau _e$ would leave the
astrophysical parameters in a reionization model as the limiting
uncertainties (Venkatesan 2002). It is important to point out that a
low $\tau _e$ does not, by itself, disfavor the VMS hypothesis. Because
their ionizing efficiency is lower than the peak (see Figure 2), a
strong-VMS model could give a low $\tau _e$ with appropriately low star
formation efficiency or photon escape fraction (if reionization is late),
or with a short time available for metal-free star formation. Unless it
settles on an extreme high value, $\tau _e$ is not likely by itself to
provide decisive constraints on the IMF.

This latter point emphasizes the critical importance of robust constraints
from nucleosynthesis as the remaining means of discriminating degenerate
IMF cases (such as a strong VMS model versus a weak VMS model without
$M < 10$ \msun, see Table 1).  At the present time, EMP stars appear
to exclude VMS. This could change because of developments in the
observations (new EMP stars with lower [Fe/H]) or refinements in the
theory (reconciliation of PISN to the Fe-peak elements would revitalize
the weak VMS hypothesis). A particular concern would be evidence that the
sample of EMP stars we presently take as the ``second generation'' is not,
in fact, the product of the first stars. Beyond their chemically primitive
atmospheres and ages, there is no hard evidence for this assumption. A
new sample of metal-poor stars with \feh\ $< -4$ could change the outcome
if they were better matched to the yields of PISN and/or clearly devoid
of \rp\ elements. There is also room for significant progress on the
theoretical front: additional physics such as asymmetry, mass loss,
and rotation could be added to the hypernovae and PISN models, perhaps
with a corresponding change in the logical basis of the VMS hypothesis.

The reionization results in Table 1 and the ionizing efficiencies in
Figure 2 suggest another observational discriminant between our
proposed first-stars IMF and one composed purely of VMS.  Although the
H~I ionizing efficiency of VMS declines from the peak at $M \sim 120$
\msun, the ionizing efficiency for He~II continues to increase into the
VMS range. This feature of VMS has two potential observational
consequences. First, VMS will be more efficient at producing the He~II
1640 \AA\ recombination line, which has been suggested as a key direct
signature of the first stars (Tumlinson, Giroux, \& Shull 2001;
Schaerer 2002). Although the many degeneracies in the age and
metallicity dependence of the $\lambda$1640 line luminosity could give
an ambiguous result, an unusually high value could be uniquely
attributed to VMS. The second possible signature is more speculative:
the He~II Gunn-Peterson test at high redshift is, in principle,
sensitive to the spectrum of the first ionizing sources.  A VMS IMF may
reionize He~II in the IGM before $z = 6$, while a non-VMS IMF may or
may not (see Table 1). Even if both IMFs can produce early He~II
reionization, they may still be discriminated by a Gunn-Peterson test.
To see why, we recall the ratio $\eta = N_{\rm HeII}/N_{\rm HI}$
defined by Fardal, Giroux, \& Shull (1998) to characterize the ionizing
sources of the IGM. Harder QSO-type spectra have $\eta \simeq 10 - 30$,
while ordinary stars produce $\eta \simeq 100 - 400$.  The increased
He~II ionizing efficiency of metal-free stars in a normal IMF can
produce $\eta \sim 30$ (Tumlinson \& Shull 2000). By excluding
inefficient low-mass stars, our proposed first-stars IMF gives $\eta
\sim 20$. A pure VMS IMF could result in $\eta \lesssim 5$. Recent work
at $z \sim 3$ with the {\it Far Ultraviolet Spectroscopic Explorer}
(Shull et al. 2004; Zheng et al. 2004) has shown that $\eta < 10$ and
$\eta > 30$ can be discriminated.  The feasibility of this test would
depend largely on the chance combination of a QSO with $z = 6 - 8$ and
little intervening absorption of the near-ultraviolet continuum by the
common Ly$\alpha$ forest or Lyman limit systems at $z = 1 - 2.5$.
Although such a coincidence is unlikely, QSOs discovered at $z = 6 - 8$
should be observed in the near UV to qualify them for this test.
Obtaining adequate S/N to accurately measure the predicted low values
of the He~II opacity would likely require a large collecting area in
the space ultraviolet, perhaps from a successor to the {\it Hubble
Space Telescope}.  If the practical problems can be overcome, the
direct detection of He~II $\lambda$1640 and the high-$z$ He~II
Gunn-Peterson test provide some means of testing the first-stars IMF.

Finally, we add that, as a theoretical point of interest, the lower
mass limit of the IMF can be just as important as the better-studied
upper limit, because it can determine the total ionizing efficiency of
the population. The lowest mass achievable by metal-free stars may
depend on the details of primordial cooling, rotation, fragmentation,
feedback, and magnetic fields in unknown ways. Attention has been
focused on modeling the formation of very massive first stars; our
results indicate that potential mechanisms for forming low-mass
metal-free stars should be pursued. This is especially true if $\tau
_e$ settles to a high value that requires high ionizing efficiency
and the present nucleosynthesis constraints on the strong VMS
hypothesis are confirmed.

Our proposed IMF can solve some key puzzles in the study of the first
stars and reionization. However, there are still significant open
questions about the formation of metal-free stars (especially the
feedback mechanisms that influence their final masses), and the unknown
time over which metal-free star formation persists.  Furthermore, there
may be no high-redshift direct tracers that can independently constrain
the proposed IMF, especially since its most salient feature is the
absence of stars with 1 -- 10~\msun\ that are not directly observable
at high-redshift. Until the launch of the {\it James Webb Space
Telescope}, which may be able to directly detect the first stars,
future work should focus on more sophisticated treatments of the
indirect tracers, particularly the EMP stars, as sensitive indicators
of the first stellar generation. Nonetheless, we have shown that VMS
are not needed for reionization if the IMF is suitably constructed, and
that such an IMF is better justified on the grounds of the nucleosynthetic
data than either version of the VMS hypothesis. In the future, better
understanding of primordial star formation, more refined models of
PISN, or a new regime of extremely metal-poor stars with more VMS-like
metal abundances may change the picture. For now, we conclude only that
our proposed IMF is better matched to the available constraints.

\acknowledgements

We acknowledge helpful discussions with Jim Truran and constructive
comments from Sergei Levshakov, Avi Loeb, and an anonymous
referee. A.~V. is supported by an NSF Astronomy and Astrophysics
Postdoctoral Fellowship under award AST-0201670. J. M. S. acknowledges
support at the Colorado astrophysical theory program from NASA LTSA
grant NAG5-7262 and NSF grant AST02-06042.

\begin{figure}
\centerline{\epsfxsize=\hsize{\epsfbox{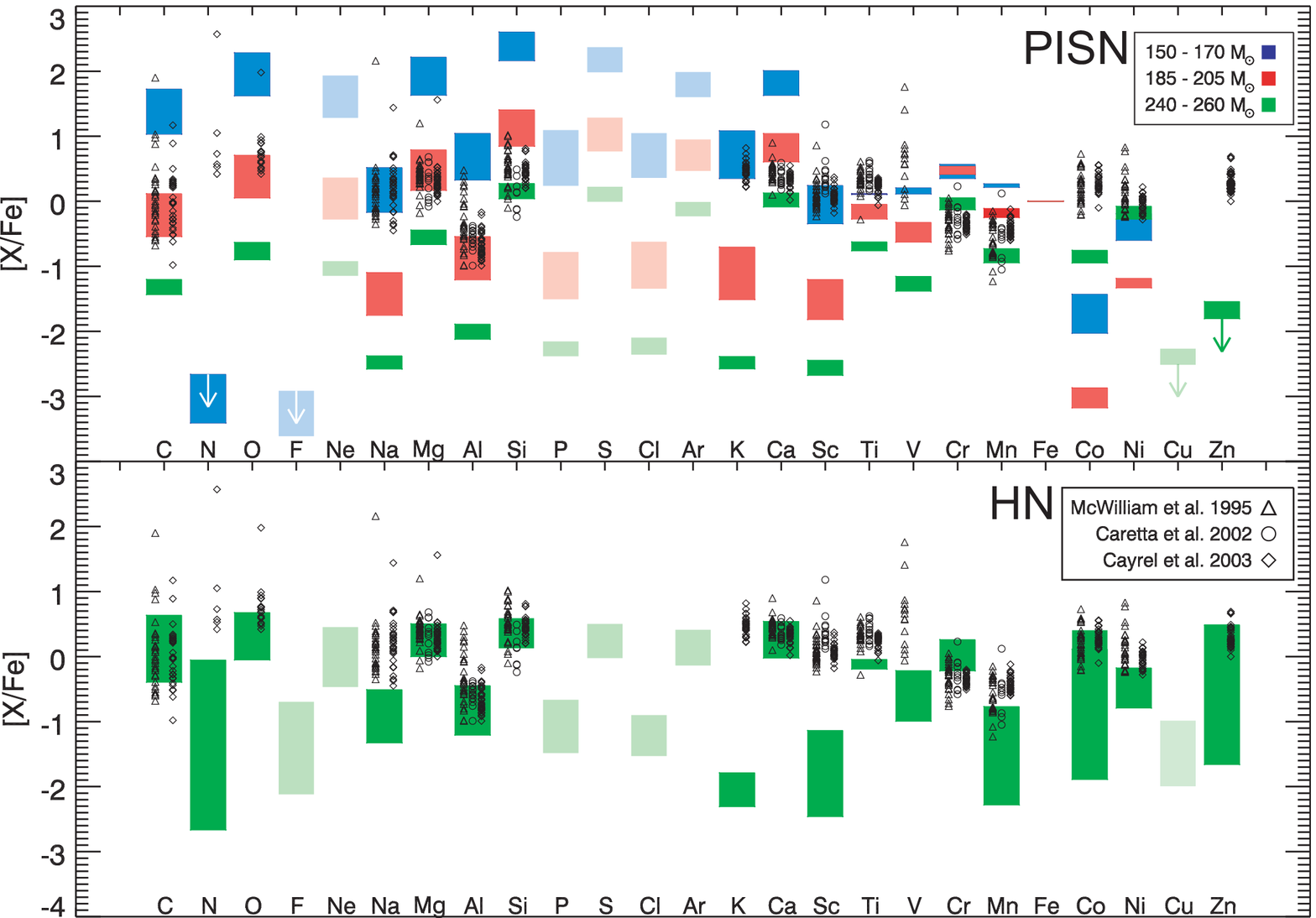}}}
\figcaption{Observed element abundances for 63 metal-poor halo stars
from McWilliam et al. (1995), Carretta et al. (2002), and Cayrel et al.
(2003). These stars range over [Fe/H] = $-2$ to $-4$. For duplicated
stars, the higher-resolution data from Cayrel et al. (2003) are used.
In the upper panel, we compare the observed abundances to the
theoretical yields for VMS pair-instability supernovae from $M = 140 -
260$~\msun. These yields are plotted separately for three ranges of
stellar mass as indicated. Clearly, VMS of a single mass cannot explain
the observed yields. Also, the extreme odd-even effect predicted for
VMS is not apparent in the data.  In the lower panel, we compare the
observed abundances to the theoretical hypernova yields of Cr, Mn, Co,
and Zn from Umeda \& Nomoto (2004), adjusted for small changes in
$Y_e$, and for all other elements from Umeda \& Nomoto (2002), over the
range $M = 1 - 50 M_\odot$ and $E_{\rm 51} = 1 - 100$. These yields
provide a superior fit to the observed patterns, particularly for the
Fe-peak elements (Cr -- Zn).}
\end{figure}

\begin{figure}
\centerline{\epsfxsize=\hsize{\epsfbox{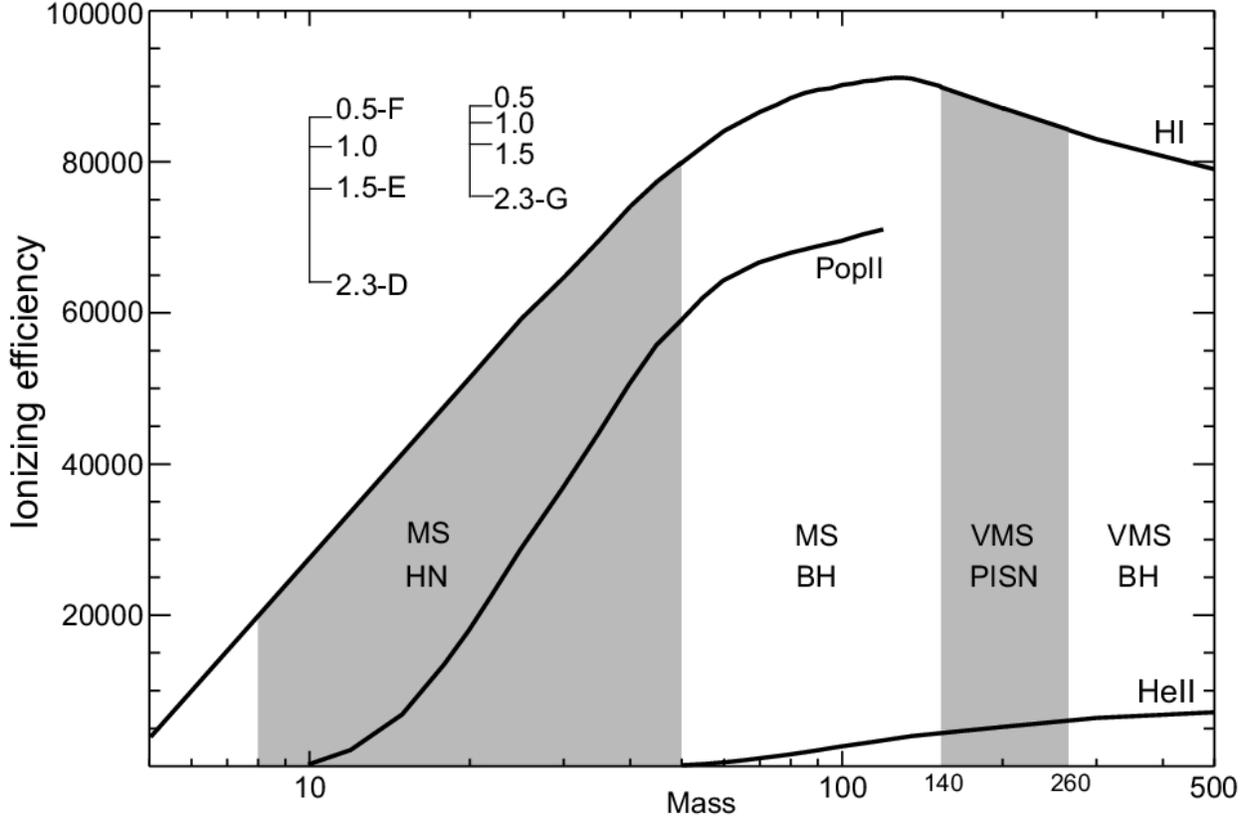}}}
\figcaption{Number of H~I and He~II ionizing photons per baryon
(``ionizing efficiency'') as a function of stellar mass, and
integrated over the star's lifetime.  Note the sharp rise from $M =
10$ to $M = 50$~\msun\ and the gradual decline for $M > 120$ \msun,
owing to asymptotic effective temperature gains and decreasing stellar
lifetimes. Also shown at 10 and 20 $M_\odot$ are the IMF-weighted values
for IMF slope between 0.5 and 2.3 and the indicated lower mass limits
(these are marked with their corresponding letters from Table 1 where
applicable).  The ionizing efficiency peaks before the VMS mass range and
decreases steadily at higher mass.  Thus, early reionization might be
achieved by a ``normal'' IMF from which the low-mass stars are absent,
rather than a VMS IMF.  We also show comparison a characteristic curve
for Population II, derived from models run with the Starburst99 code
(Leitherer et al. 1999) and assuming standard mass loss and $Z = 0.001$.
\label{phot}} \vspace{0.2in} \end{figure}

\pagebreak
\renewcommand{\arraystretch}{1.2}
\begin{table}[ht]
\begin{center}
\caption{Effects of Stellar IMFs on Reionization}
\vspace{0.1in}
\begin{tabular}{ccccrrccrrc}
\tableline \tableline
Model & IMF  & $Q_{\rm HI}$  & $Q_{\rm HeII}$  & \multicolumn{3}{c}{$10^7$ yr (intrahalo)}  & &
\multicolumn{3}{c}{$10^8$ yr (halo-halo)}  \\ \tableline
 & &  & &  $z_{\rm r, H}$ & $z_{\rm r, He}$  & $\tau_e$ & & $z_{\rm r, H}$ &
$z_{\rm r, He}$  & $\tau_e$   \\ \tableline
A & 10$^2$--10$^3$~\msun\ & 1.6(48) & 3.8(47) & 13.4 & 10.3 & 0.134 & & 15.4 & 13.8 &
0.157 \\
B & 1--100~\msun\ & 1.1(47) & 1.5(45) & 11.4 & 0.5 & 0.095 & & 11.8 & 1.6 & 0.102  \\
C & 1--140~\msun\ & 1.5(47) & 3.3(45) & 11.5 & 0.9 & 0.097 & & 12.1 & 3.9 & 0.109 \\
D & 10--140~\msun\ & 4.3(47) & 9.8(45) & 12.0 & 1.6 & 0.105 & &  13.7 & 7.0 & 0.128  \\
E & 10--140~\msun, $\alpha$=1.5 & 6.7(47) & 2.2(46) & 12.3 & 3.3 & 0.112 &
& 14.3 & 8.9 & 0.137  \\
F & 10--140~\msun, $\alpha$=0.5 & 8.9(47) & 3.8(46) & 12.6 & 4.4 & 0.117 &
& 14.7 & 10.0 & 0.143  \\
G & 20--140~\msun\ & 6.1(47) & 1.5(46) & 12.3 & 2.5 & 0.110 & & 14.2 & 8.0 & 0.134
\end{tabular}
\vspace{-0.3in} \tablecomments{This table shows the IMFs considered in
this work and the VMS IMF from Bromm, Kudritzki, \& Loeb (2001), with
the mass-normalized ionizing photon rates $Q_i$ (photons $s^{-1}$
\msun$^{-1}$) for H~I and He~II at a starburst age of 1 Myr. In our
notation 1(50) $= 1 \times 10^{50}$. The reionization redshifts, $z_r$
for H~I and He~II are also shown for each case, and the calculated
electron-scattering optical depth $\tau _e$ (including the contribution
from He II). We show results for two metal-free star formation cutoff times of $10^7$
yr and $10^8$ yr, corresponding to the intra-halo and halo-halo metal
enrichment timescales, respectively (see \S~4.2). All cases are for
metal-free stars and an IMF given by $dN/dM \propto M^{-\alpha}$, with
$\alpha$ = 2.35 unless otherwise stated.}
\end{center}
\end{table}


\begin{references}
\reference{c02} Abel, T., Bryan, G., Norman, M. L. 2000, Science, 293, 93
\reference{wa02} Aoki, W., et al. 2003, \apj, 576, L141
\reference{rb01} Becker, R., et al. 2001, AJ, 122, 2850
\reference{b01b} Bromm, V., Kudritzki, R.-P., \& Loeb, A. 2001, \apj, 552, 464
\reference{b03b} Bromm, V., Yoshida, N., \& Hernquist, L. 2003, \apj, 596, 137
\reference{b03a} Bromm, V., \& Loeb, A. 2003, Nature, 425, 812
\reference{b03a} Bromm, V., \& Loeb, A. 2004, New Astronomy, 9, 353 
\reference{burr00} Burris, D., et al. 2000, \apj, 544, 203
\reference{cay03} Cayrel, R., et al. 2003, A\&A, 416, 1117 
\reference{c02} Carretta, E., Gratton, R., Cohen, J.G., Beers, T. C.,
        \& Christlieb, N. 2002, \aj, 124, 481
\reference{he0107} Christlieb, N., et al. 2002, Nature, 419, 904
\reference{cfw} Ciardi, B., Ferrara, A., \& White, S. D. M. 2003, MNRAS, 344, 7
\reference{Cen03a} Cen, R. 2003a, \apj, 591, 12
\reference{Cen03b}Cen, R. 2003b, \apj, 591, L5
\reference{d} Depagne, E., et al. 2002, A\&A, 390, 187
\reference{sd01} Djorgovski, S. G., Castro, S. M., Stern, D., \&
        Mahabal, A. 2001, \apj, 560, L5
\reference{fgs98} Fardal, M., Giroux, M. L., \& Shull, J. M. 1998, AJ, 115, 2206
\reference{ftc02} Fields, B. D., Truran, J. W., \& Cowan, J. J. 2002, 575, 845
\reference{fry} Fryer, C. L., Woosley, S. E., \& Heger, A. 2001, 550, 372
\reference{fk} Fukugita, M., \& Kawasaki, M. 2003, MNRAS, 343, 25
\reference{gia} Giavalisco, M., et al. 2004, \apj, 600, L103
\reference{go}  Gnedin, N. Y., \& Ostriker, J. P. 1997, 486, 581
\reference{g00} Gnedin, N. Y. 2000, \apj, 535, 530
\reference{hh03} Haiman, Z., \& Holder, G. P. 2003, 595, 1
\reference{mh03} Hamuy, M. 2003, \apj, 582, 905
\reference{HW02} Heger, A., \& Woosley, S. E. 2002, \apj, 567, 532 (HW02)
\reference{hh} Hui, L., \& Haiman, Z. 2003, ApJ, 596, 9
\reference{kap} Kaplinghat, M., et al. 2003, ApJ,  583, 24
\reference{ak03} Kogut, A., et al. 2003, \apjs, 148, 161
\reference{s99} Leitherer et al., 1999, \apjs, 123, 3 
\reference{lev} Levshakov, S., et al. 2003, A\&A, 397, 851 
\reference{lim} Limongi, M., Chieffi, A., \& Bonifacio, P. 2003, \apj, 594, L123
\reference{mar} Marigo, P., Girardi, L., Chiosi, C., \& Wood, P. R.
        2001, A\&A, 371, 152
\reference{mcw95} McWilliam, A., Preston, G. W., Sneden, C., \& Searle,
        L. 1995, \aj, 109, 2757
\reference{mcw97} McWilliam, A. 1997, ARA\&A, 35, 503
\reference{jme03} Miralda-Escud\'e, J. 2003, \apj, 597, 66
\reference{nom03} Nomoto, K., et al. 2003, in Carnegie Observatories
        Astrophysics Series, Vol. 4: Origin and Evolution
        of the Elements, ed. A. McWilliam and M. Rauch, 1
\reference{EMP9} Norris, J. E., et al. 2002, \apj, 569, L107
\reference{oh01} Oh, S. P., Nollett, K. M., Madau, P., \&
   Wasserburg, G. J. 2001, \apj, 562, L1
\reference{ko00} Omukai, K. 2000, \apj, 534, 809
\reference{OP} Omukai, K., \& Palla, F. 2002, ApJ, 589, 677
\reference{jpo81} Ostriker, J. P., \& Cowie, L. L. 1981, \apj, 243, L127
\reference{pw02} Prochaska, J. X., \& Wolfe, A. M. 2002, 566, 68
\reference{QW02} Qian, Y.-Z., \& Wasserburg, G. J. 2002, \apj, 567, 515 (QW02)
\reference{rgs1} Ricotti, M., Gnedin, N. Y., \& Shull, J. M. 2002a, \apj, 575, 33 (RGS)
\reference{rgs2} Ricotti, M., Gnedin, N. Y., \& Shull, J. M. 2002b, \apj, 575, 49
\reference{es02} Scannapieco, E., Ferrara, A., \& Madau, P. 2002, \apj, 574, 590
\reference{es03} Scannapieco, E., Schneider, R., \& Ferrara, A. 2003, \apj, 589, 35
\reference{sch} Schaerer, D. 2002, A\&A, 382, 28
\reference{rs02} Schneider, R., Ferrara, A., Natarajan, P., \& Omukai, K. 2002, \apj, 571, 30
\reference{jms81} Shull, J. M., \& Silk, J. 1981, \apj, 249, 26
\reference{jms04} Shull, J. M., Tumlinson, J., Giroux, M. L., Kriss, G. A., \& Reimers, D.
          2004, \apj, 600, 570 
\reference{som} Somerville, R. S., Bullock, J. S., \& Livio, M.  2003, ApJ, 593, 616
\reference{ds03} Spergel, D., et al. 2003, \apjs, 148, 175
\reference{ccs99} Steidel, C. C., Adelberger, K. L., Giavalisco, M., Dickinson, M.,
                       \& Pettini, M. 1999, \apj, 519, 1
\reference{Tm02} Tan, J., \& McKee, C. M. 2002, in ``The Emergence of Cosmic Structure'',
                Proceedings of the AIP, 666, 93
\reference{Tm02} Tan, J., \& McKee, C. M. 2004, astroph/0307414
\reference{T03} Tegmark, M. 2004, ApJ, in press, astro-ph/0310723
\reference{jwt02} Truran, J. W., Cowan, J. J., Pilachowski, C. A., \&
      Sneden, C. 2002, \pasp, 114, 1293
\reference{ts00} Tumlinson, J., \& Shull, J. M.. 2000, \apj, 528, L65
\reference{tgs01} Tumlinson, J., Giroux, M. L., \& Shull, J. M. 2001, \apj, 550, L1
\reference{tsv02} Tumlinson, J., Shull, J. M., \& Venkatesan, A. 2003, \apj, 584, 608 (TSV03)
\reference{un02} Umeda, H., \& Nomoto, K. 2002, \apj, 565, 385
\reference{un02} Umeda, H., \& Nomoto, K. 2003, Nature, 422, 871
\reference{un03} Umeda, H., \& Nomoto, K. 2004, astro-ph/0308029 (UN04)
\reference{v02} Venkatesan, A. 2002, \apj, 572, 15
\reference{tsv02} Venkatesan, A., Tumlinson, J., \& Shull, J. M. 2003, \apj, 584, 621 (VTS03)
\reference{vt03} Venkatesan, A., \& Truran, J. W. 2003, ApJ, 594, L1
\reference{vsf04} Venkatesan, A., Schneider, R., \& Ferrara, A. 2004, MNRAS, 349, L43 
\reference{w03} Wada, K., \& Venkatesan, A. 2003, \apj, 591, 38
\reference{WQ00} Wasserburg, G. J., \& Qian, Y.-Z. 2000, \apj, 529, L21 (WQ00)
\reference{W94} Woosley, S. E., et al. 1994, \apj, 433, 229
\reference{wl03} Wyithe, S., \& Loeb, A. 2003, \apj, 586, 693 
\reference{Z04} Zheng, W., et al. 2004, \apj, 605, 671 
\end{references}
\end{document}